\documentclass[showpacs,preprintnumbers,amsmath,amssymb,superscriptaddress,nofootinbib,english]{revtex4}

\usepackage{graphicx}
\usepackage{dcolumn}
\usepackage{bm}
\usepackage{epsfig}
\usepackage{graphicx}
\usepackage{hyperref}
\usepackage[usenames]{color}
\usepackage{url}

\hypersetup{
    colorlinks=true,
    linkcolor=green,
    citecolor=blue,
}

\newcommand{\remove}[1]{}

\def\be{\begin{equation}}
\def\ee{\end{equation}}
\def\ba{\begin{eqnarray}}
\def\ea{\end{eqnarray}}

\begin{document}

\title{Supersymmetron}

\author{Philippe~Brax}
\email[Email address: ]{philippe.brax@cea.fr}
\affiliation{Institut de Physique Theorique, CEA, IPhT, CNRS, URA 2306, F-91191Gif/Yvette Cedex, France}

\author{Anne-Christine~Davis}
\email[Email address: ]{a.c.davis@damtp.cam.ac.uk}
\affiliation{DAMTP, Centre for Mathematical Sciences, University of Cambridge, Wilberforce Road, Cambridge CB3 0WA, UK}

\date{\today}

\begin{abstract}
We consider a supersymmetric model of dark energy coupled to cold dark matter: the supersymmetron. In the absence of cold dark matter, the supersymmetron converges to a supersymmetric
minimum with a vanishing cosmological constant. When cold dark matter is present, the supersymmetron evolves to a matter dependent minimum where its energy density does not vanish. In the early universe until the recent past of the Universe, the energy density of the supersymmetron is negligible compared to the cold dark matter energy density.  Away from the supersymmetric minimum,  the equation of state of the supersymmetron is constant and negative.
When the supersymmetron reaches the neighbourhood of the supersymmetric minimum, its equation of state vanishes rapidly. This leads to an acceleration of the universe which is transient unless supersymmetry breaking induces  a pure cosmological constant and acceleration of the Universe does not end. Moreover, we find that the  mass of supersymmetron is always greater than the gravitino mass. As a result, the supersymmetron generates a short ranged fifth force which evades gravitational tests. On the other hand, we find that
the supersymmetron may lead to relevant effects on large scale structures.

\end{abstract}

\maketitle

\section{Introduction}

Dark energy models of cosmic acceleration suffer from many problems\cite{Copeland:2006wr}. At best they can be treated as low energy field theories valid at energies well below the electron mass, corresponding to the very late phase of the Universe. Hence these models need to be embedded in a better defined theory whose ultra violet behaviour is  under control. So far, no such complete scenario has been constructed. Dark energy models also seem to require the existence of a very light scalar field whose coupling to matter leads to a long ranged fifth force whose presence is at odds with current gravitational tests, see  \cite{Jain:2010ka} and references therein. Screening mechanisms have been invoked in order to alleviate this problem \cite{Khoury:2003aq,Khoury:2003rn,Brax:2004qh,Mota:2006ed,Mota:2006fz,Hinterbichler:2010es,Brax:2010gi,Nicolis:2008,Olive:2007aj}.
Axion-like particles with derivative couplings to matter are also possible candidates\cite{Choi:1999xn}.

In this letter, we will revisit some of these issues in a very simple context. Indeed, it has long been advocated that a natural solution to the cosmological constant problem could involve supersymmetry \cite{Brax:1999gp,Brax:1999yv,Copeland:2000vh}.
As a matter of fact, globally supersymmetric theories are protected by a non-renormalisation theorem and possess vacua with vanishing energy densities \cite{Seiberg:1993vc}. Of course, the  absence of experimental evidence in favour of supersymmetric partners implies that supersymmetry must be broken in the matter sector comprising the particles of the standard model. As such radiative corrections  in the softly broken Minimal Supersymmetric Standard Model (MSSM) give a contribution to the vacuum energy which is already far larger than the critical density of the Universe. Unfortunately, we will have nothing new  to say about this thorny question. We will simply assume that an unknown mechanism specific to the matter sector allows for a  cancellation of all the matter contributions to the vacuum energy density.

On the other hand, it could well be that the dark sector of the Universe, composed of the still undiscovered Cold Dark Matter (CDM) and Dark Energy (DE), could be described by a globally supersymmetric theory. In such a case the vanishingly small amount of dark energy which is necessary to generate the acceleration of the Universe could result from a small cosmological breaking of supersymmetry due to the non-zero CDM energy density. Such a scenario would naturally lead to a close relationship between the dark energy and the CDM energy densities. Of course, one should also make sure that corrections to the globally supersymmetric scalar potential coming from the soft supersymmetry breaking in the MSSM sector  do not spoil the CDM-DE correspondence and the properties of the scalar potential in the late time Universe. Moreover, if the dark sector couples to ordinary matter, it should be ascertained   that no long range fifth force is  present in this scenario.

In the following, we will present a model of dark energy coupled to dark matter in a globally supersymmetric context. Both the superpotential and the K\"ahler potential have non-renormalisable terms suppressed by a very large scale. When CDM is present, we find that supersymmetry is cosmologically broken with a dynamical minimum which is an attractor. In the future of the  Universe, this minimum becomes close to the supersymmetric minimum where no acceleration is present. In the recent past, the Universe accelerates with a constant equation of state.   Moreover we find that the supersymmetry breaking corrections are suppressed and the dark energy field acquires a mass larger than the gravitino mass thus evading local gravitational tests.  Due to its supersymmetric properties and its link to the cosmological breaking of supersymmetry, we have called the dark energy field in this model "supersymmetron".

The supersymmetron allows one to have a description of dark energy and the acceleration of the Universe with no need for a cosmological constant which  could
spring from the vacuum energy in the supersymmetry breaking sector. In fact we find that the supersymmetron dark energy can at most account for fifty per cent of the
critical density of the Universe. In view of the stronger and stronger observational evidence that dark energy should represent almost three quarters of the energy budget
of the Universe\cite{Perlmutter:1998np,Riess:1998cb}, we must introduce a  cosmological constant. This woeful nuisance is intriguing and getting rid of this  contribution to the vacuum energy is left for future work. On the positive side,
it turns out that the symmetron model potentially modifies the growth of cosmological structures in a novel way. The investigation of this issue is under way.

In the next section, we will present the supersymmetron model and investigate its properties. The cosmological evolution and consequences are studied in the
following section. Our conclusions are presented in section $4$.

\section{Supersymmetron Dark Energy}

We are interested in supersymmetric models of dark energy coupled to Cold Dark Matter (CDM). As long as the CDM energy density vanishes, the theory has a supersymmetric minimum with a vanishing vacuum energy. When CDM develops a non-vanishing energy density, the supersymmetric minimum is lifted and the residual amount of vacuum energy density tracks the CDM background density. Eventually, the CDM matter density goes to zero and the vacuum converges to the true supersymmetric vacuum.

We assume that the models under study are low energy effective theories valid after Big Bang Nucleosynthesis (BBN) at energies well below the electron mass threshold. We will describe theories of the dark sector of the Universe encompassing both dark energy and dark matter. In this dark sector we posit that the Universe is described by a globally supersymmetric theory. If supersymmetry is also present in the standard model sector, it has to be broken in a way which is either not communicated to the dark sector, or very weakly in order to essentially preserve the properties of global supersymmetry.

\subsection{Acceleration and scalar-tensor theories}

Scalar-tensor theories appear naturally in cosmology when describing dark energy\cite{Khoury:2003rn,Brax:2004qh,Mota:2006ed,Brax:2010gi,Olive:2007aj,Hinterbichler:2010es,Amendola:1999er,Anderson:1997un,Farrar:2003uw,Bean:2008ac,Brookfield:2007au,LaVacca:2009yp}. We will find that the supersymmetron can be seen as a particular scalar-tensor model.
Let us consider the theory defined by the following Lagrangian involving both gravity and the dark energy field $\phi$
\begin{equation}
S=\int d^4x\sqrt{-g}\left\{\frac{m_{Pl}^2}{2}{\cal
R}-\frac{k^2(\phi)}{2}(\partial\phi)^2- V(\phi)\right\} - \int d^4x{\cal
L}_m(\psi_m^{(i)},\tilde g_{\mu\nu})\,, \label{action}
\end{equation}
where ${\cal L}_m$ is the matter Lagrangian and the minimally coupled metric is
\begin{equation}
\tilde g_{\mu\nu}= A^2(\phi) g_{\mu\nu}.
\end{equation}
The matter Lagrangian contains  particles such as Cold Dark Matter (CDM) particles.
To study the cosmology of the model, we only need the Friedmann equation and the conservation of the Cold Dark Matter density
\begin{equation}
H^2= \frac{\rho_T}{3m_{\rm Pl}^2}
\end{equation}
where $\rho_T$ is the total energy density. The CDM energy density $\rho_\psi$ is conserved implying that
\begin{equation}
\dot \rho_\psi =-3H \rho_\psi.
\end{equation}
The energy density in the Einstein frame defined by the action (\ref{action}) is not conserved and is related to the conserved matter density as
$\rho_E= A(\phi) \rho_\psi$.
The dynamics of the scalar field can be deduced from the Klein-Gordon equation
\begin{equation}
\ddot\varphi +3H\dot\varphi = -\frac{dV_{\rm eff}}{d\varphi}
\end{equation}
where we take into account the possibility that the scalar field $\phi$ is not normalised and define the normalised field as
\begin{equation}
d\varphi\equiv k(\phi) d\phi.
\end{equation}
The effective potential is defined as
\begin{equation}
V_{\rm eff}(\phi)= V(\phi) + (A(\phi)-1)\rho_\psi.
\end{equation}
Because of the interaction between the scalar field and CDM, the energy momentum tensor of the scalar field is not conserved. Only the total energy momentum is conserved
\begin{equation}
\dot \rho_T = -3H (\rho_T + p_T)
\end{equation}
where the total energy density is
\begin{equation}
\rho_T\equiv \rho_\psi + \rho_\phi
\end{equation}
with
\begin{equation}
\rho_\phi= \frac{\dot \varphi^2}{2} + V_{\rm eff}, \ p_T\equiv p_\phi= \frac{\dot \varphi^2}{2} - V.
\end{equation}
It is crucial to notice that the energy density of the scalar field involves the effective potential $V_{\rm eff}$ while the pressure only involves $V_F$. This is a crucial feature of scalar-tensor theories.

We can define the effective
 equation of state of the dark energy fluid as
\begin{equation}
w_{\phi}= \frac{p_\phi}{\rho_\phi}\approx -\frac{V(\phi)}{V_{\rm eff}(\phi)},
\end{equation}
the latter when the kinetic energy is negligible. The evolution of the scalar field energy density satisfies
\begin{equation}
\frac{d\rho_\phi}{dt}= -3 H(1+ \omega_\phi) \rho_\phi.
\end{equation}
Using the Friedmann equation we find that
\begin{equation}
\dot H= -\frac{\dot \varphi^2+ \rho_\psi A(\phi)}{2}
\end{equation}
and therefore the Raychaudhuri equation
\begin{equation}
\frac{\ddot a}{a} = -\frac{2\dot\varphi^2 -{\rho_\psi A(\phi)} + 2{V}}{6m^2_{\rm Pl}}
\end{equation}
which is identical to the usual Raychaudhuri equation involving the effective equation of state $w_\phi$
\begin{equation}
\frac{\ddot a}{a} = -\frac{1}{6m_{\rm Pl}^2} ( \rho_\psi + (1+3w_{\phi}) \rho_{\phi})\equiv -\frac{1}{6m_{\rm Pl}^2} (1+3w_{T}) \rho_{T}
\end{equation}
where we have defined the total equation of state $w_T= \frac{p_T}{\rho_T}$
The universe is accelerating  provided $\ddot a \ge 0$ which leads to
\begin{equation}
w_T\le-\frac{1}{3}
\end{equation}
which is equivalent to
\begin{equation}
w_\phi\le -\frac{1}{3}(1+ \frac{\rho_\psi}{\rho_\phi}).
\end{equation}
This reduces to
\begin{equation}
V(\phi) \ge \frac{1}{2} \rho_\psi A(\phi)
\end{equation}
as long as the kinetic energy is negligible.
Acceleration of the Universe is also related to the acceleration parameter
\begin{equation}
q=-\frac{\ddot a a}{\dot a^2}= \frac{1}{2} (1+ 3w_T)
\end{equation}
which must be negative. Supernovae data constrain this parameter \cite{Perlmutter:1998np, Riess:1998cb}. In practice, we will use
\begin{equation}
q=\frac{1}{2}\frac{ \rho_\psi + (1+3w_{\phi}) \rho_{\phi}}{\rho_\psi +\rho_\phi}
\end{equation}
and apply it to cases where the kinetic energy of the scalar field is negligible.

Bounds on the acceleration parameter are related to the behaviour of the equation of state $w_T$.  When the scalar field and matter are not coupled, the scalar field satisfies the Null Energy Condition (NEC) $w_\phi \ge -1$. Similarly the overall two-component system, i.e. scalar and matter, satisfies the NEC $w_T\ge -1$ too. The former inequality does not hold anymore when a coupling is present. Indeed, it is violated when
$\dot\varphi^2 +(A(\phi)-1)\rho_\psi <0$ which can happen when $A<1$ and the kinetic energy is negligible\cite{astro-ph/0510628}. On the other hand the total equation of state always  satisfies the NEC and therefore $q\ge -1$.
\subsection{The supersymmetron}

In globally supersymmetric models of  the scalar sector, models are specified by their K\"ahler potential and the superpotential. We choose the superpotential to be
\begin{equation}
W= g\phi \phi_-\phi_+ + m\phi_- \phi_+ + \frac{\phi^\alpha}{\Lambda_0^{\alpha-3}} + \frac{\phi^\beta}{\Lambda_2^{\beta-3}}
\end{equation}
and the K\"ahler potential
\begin{equation}
K= \vert \phi_-\vert^2 + \vert \phi_+\vert^2 + \frac{\vert \phi\vert ^{2\beta}}{\Lambda_1^{2\beta-2}}
\end{equation}
where $\phi$ is the dark energy scalar superfield and $\phi_\pm$ the CDM particles. The bosonic part of the dark energy superfield describes dark energy per se. Notice that the superpotential and the K\"ahler potential have non renormalisable terms which should be deduced from the high energy dynamics of the model, possibly after integration over extra massive degrees of freedom. The scalar potential is given by the F-term $V_F= K^{\phi\bar\phi} \vert\partial_\phi W\vert^2$ which reads:
\begin{equation}
V_F= \vert \Lambda^2 + \frac{M^{2+n/2}}{\phi^{n/2}}\vert^2.
\label{pot}
\end{equation}
Putting $\phi=\vert \phi \vert e^{i\theta}$,  the potential is minimised when $\cos (\alpha -\beta)\theta =-1$. The angular field is stabilised at this minimum with a mass which is always much greater
than the gravitino mass (see later) implying that
\begin{equation}
V_F= ( \Lambda^2 - \frac{M^{2+n/2}}{\vert \phi\vert ^{n/2}})^2.
\label{poti}
\end{equation}
We have identified
\begin{equation}
\Lambda^{4}= \frac{\Lambda_1^{2\beta-2}}{\Lambda_2^{2\beta -6}}, \ \ M^{4+n}= \frac{\alpha^2}{\beta^2} \frac{\Lambda_1^{2\beta-2}}{\Lambda_0^{2\alpha -6}}
\end{equation}
and the index
\begin{equation}
n= 2(\beta -\alpha).
\end{equation}
Notice that when $\vert \phi\vert $ tends to zero, the potential is of the inverse power law type as in the Ratra-Peebles model\cite{Ratra:1987rm}. Moreover the potential becomes equivalent to a pure cosmological constant $\Lambda^4$ for large values of the field $\vert \phi\vert$.
In the following we always simplify the notation and write  $\phi\equiv \vert \phi\vert$.

The potential has a supersymmetric minimum with vanishing energy for
\begin{equation}
\phi_{\rm min}= (\frac{\beta}{\alpha} \frac{\Lambda_0^{\alpha-3}}{\Lambda_2^{\beta -3}})^{1/(\alpha-\beta)}
\end{equation}
where $\partial_\phi W=0$.
This supersymmetric minimum will be lifted when CDM develops a non-zero number density. Indeed
the mass term Lagrangian for the fermionic CDM particles is
\begin{equation}
{\cal L}_f= m(1+ \frac{g}{m} \phi) \psi_+ \psi_-
\end{equation}
corresponding to a scalar field dependent mass.
Cosmologically the CDM particles develop a non-vanishing number density $n_\psi= <\psi_+ \psi_->$ early in the universe leading to a new contribution
to the scalar potential
\begin{equation}
V_{\rm eff}(\phi)= V_F(\phi) + \frac{g}{m} \phi \rho_\psi
\end{equation}
where
\begin{equation}
\rho_\psi= m n_\psi
\end{equation}
is the CDM energy density.
The fermion masses and the effective potential  correspond to a  coupling of the dark energy particle to CDM which is identified as
\begin{equation}
A(\phi)= 1+ \frac{g}{m}{\phi}
\end{equation}
when viewing this effective model as a scalar tensor theory where  CDM particles follow geodesics of the rescaled metric $\tilde g_{\mu\nu}= A^2 g_{\mu\nu}$.
The non-vanishing value of $n_\psi$  is a cosmological breaking of supersymmetry reappearing as a  Lorentz invariance  breaking term in $V_{\rm eff}\equiv V_F(\phi) + (A(\phi)-1) \rho$ depending on the CDM number density $n_\psi$.

The field $\phi$ is not canonically normalised as its kinetic term reads
\begin{equation}
{\cal L}_{\rm kin} = \frac{\beta^2 \phi^{2\beta -2}}{\Lambda_1^{2\beta -2}} (\partial \phi)^2.
\end{equation}
The normalised field is
\begin{equation}
d\varphi= \sqrt{2} \frac{\beta \phi^{\beta-1}}{\Lambda_1^{\beta-1}} d\phi\equiv k(\phi) d\phi
\end{equation}
and varies as $\phi^\beta$.
We will use both variables.
For instance we have for the mass of the dark energy particle
\begin{equation}
\frac{d^2 V_{\rm eff}}{d\varphi^2}= \frac{1}{ k^2(\phi)} \frac{d^2V_{\rm eff}}{d\phi^2} - \frac{1}{k^3(\phi)} \frac{dk}{d\phi} \frac{dV_{\rm eff}}{d\phi}
\end{equation}
and therefore  at the minimum of the effective potential, the mass of the scalar field  is
\begin{equation}
m^2_\rho= \frac{1}{ k^2(\phi_\rho)} \frac{d^2V_{\rm eff}}{d\phi^2}\vert_{\phi_\rho}
\end{equation}
where $\phi_\rho$ is the minimum of the effective potential.

\subsection{The effective vacuum}

When $\rho_\psi$ is small enough we can
expand the effective potential to second order around the supersymmetric minimum
\begin{equation}
V_{\rm eff}(\phi) \sim \frac{m_0^2}{2} (\phi-\phi_{\rm min})^2 + \frac{g}{m} \phi \rho_\psi
\end{equation}
and  find that the supersymmetric minimum is lifted by the condensate and becomes
\begin{equation}
\phi_\rho= \phi_{\rm min} -\frac{g\rho_\psi}{m m_0^2}
\label{quad}
\end{equation}
with a value of the potential at the minimum
\begin{equation}
V_{\rm eff}(\phi_\rho)= \frac{g}{m} \phi_{\rm min} \rho_\psi\equiv x \rho_\psi
\end{equation}
provided $g\rho_\psi\ll mm_0^2 \phi_{\rm min}$.
Notice that this is a small deformation of the supersymmetric vacuum with vanishing energy density. The fact that the value of the potential at the minimum
is proportional to $\rho_\psi$ comes from the fact that we are explicitly breaking supersymmetry with a non-vanishing condensate $\rho_\psi= m n_\psi$.
This is true as long as the minimum is in the vicinity of the supersymmetric minimum and the kinetic energy is negligible. The kinetic energy is largely suppressed by the $k(\phi_\rho)^2$ factor compared to the potential energy and we can safely neglect it.

When the energy density of the universe is much higher, the potential has a minimum to the left of the supersymmetric minimum on the branch
where the $M^{4+n}/\phi^n$ term dominates in $V_F$. The effective potential has a minimum with
\begin{equation}
 \phi_\rho= (\frac{nM^{4+n}m}{g \rho_\psi})^{1/(n+1)}
\label{phi}
\end{equation}
The minimum of the potential is an attractor as long as $m_\rho\gg H$. Denoting by $\rho_\infty\le \Omega_m^0 \rho_c$, the energy density when the dark energy field reaches the supersymmetric minimum, where ${\Omega_m}_0$ is the mass density
fraction today and $\rho_c$ the closure energy density, we have
\begin{equation}
 \rho_\infty\approx \frac{ nM^{4+n} m}{g \phi_{\rm min}^{n+1}}.
\label{rho}
\end{equation}
We obtain that
\begin{equation}
\phi_\rho\approx \phi_{\rm min}(\frac{ \rho_\infty}{\rho_\psi})^{1/(n+1)}.
\label{min}
\end{equation}
The mass at the minimum is given by
\begin{equation}
m^2_\rho= \frac{n(n+1) M^{4+n}}{\phi_\rho^{n+2} k(\phi_\rho)^2}
\end{equation}
implying that
\begin{equation}
\frac{m^2_\rho}{H^2}\approx x^{1-2\beta} g^{2\beta}(\frac{\Lambda_1}{m})^{2(\beta -1)} \frac{m_{\rm Pl}^2}{m^2}(\frac{\rho}{\rho_\infty})^{(2\beta-1)/(n+1)}.
\end{equation}
As long as $m\le \Lambda_1$, this is much larger than one and the minimum is an attractor.

\subsection{Mass scales}

Using  (\ref{phi},\ref{rho}), we find relations between the scales:
\begin{equation}
\Lambda_0^{\alpha -3} \approx \frac{\alpha}{\beta}\sqrt{n}  g^{n/2}\frac{\Lambda_1^{\beta -1}}{m^{n/2} \sqrt{\rho_\infty }}x ^{-(n+1)/2}
\end{equation}
and
\begin{equation}
\Lambda_2^{\beta -3}\approx \sqrt{n} \frac{\Lambda_1^{\beta -1}}{\sqrt{\rho_\infty}}x^{-1/2}
\end{equation}
from which we deduce that
\begin{equation}
\Lambda^4 \approx \frac{x\rho_\infty}{n}.
\end{equation}
Close to the supersymmetric minimum the mass is
\begin{equation}
m^2_{\rho 0}\equiv m_0^2 k(\phi_\rho)^{-2} \approx 2 (\alpha-\beta)^2 k(\phi_{\min})^{-2}\phi_{\rm min}^{-2} \frac{\Lambda_1^{2\beta -2}}{\Lambda_2^{2\beta -6}}
\end{equation}
where
\begin{equation}
m_0^2 \approx \frac{n}{2} g^2 \frac {\rho_\infty }{x m^2}
\end{equation}
which implies that
\begin{equation}
m_{\rho 0} \approx x^{1-2\beta} g^{2\beta} \frac{\sqrt{\rho_\infty}}{m} (\frac{\Lambda_1}{m})^{\beta -1}.
\label{mass}
\end{equation}
The mass on the attractor away from the supersymmetric minimum scales like
\begin{equation}
\frac{m^2_\rho}{m^2_{\rho 0}}\sim (\frac{\rho_\psi}{\rho_\infty})^{1+ \frac{(2\beta-1)}{n+1}}
\end{equation}
implying that the mass is always larger than the mass close to the supersymmetric  minimum of the potential.

\subsection{Supersymmetry breaking}

The mass at the supersymmetric  minimum is naturally large. This follows from a condition on the irrelevance of soft supersymmetry breaking terms to the dark energy potential. We assume that supersymmetry is broken at the supergravity level in a spontaneous way.
Throughout the late history of the Universe, we have that $K(\phi_\rho,\bar \phi_\rho) \ll m_{\rm Pl}^2$ implying that the main sources of corrections to the scalar potential come from
\begin{equation}
e^{K/m_{\rm Pl}^2} m_{3/2}^2 m_{\rm Pl}^2 \supset K(\phi_\rho,\bar \phi_\rho)m_{3/2}^2
\end{equation}
together with
\begin{equation}
K^{\phi\bar\phi} \vert D_\phi W\vert^2 \supset \frac{\vert K_\phi\vert^2 m_{3/2}^2}{k^2(\phi)}.
\end{equation}
All these corrections lead to a contribution to the scalar potential
\begin{equation}
\delta V \sim \frac{m_{3/2}^2 \phi^{2\beta}}{\Lambda_1^{2\beta -2}}.
\end{equation}
This increases with $\phi$ and does not modify the cosmological dynamics provided $\delta V (\phi_{\rm min})\ll \rho_\infty$ which leads to
\begin{equation}
\Lambda_1^{\beta -1} \gg \frac{m_{3/2} m^\beta}{\sqrt{\rho_\infty}}.
\end{equation}
This implies that the supersymmetry breaking correction is always negligible when the dark energy field is smaller than the field value at the
supersymmetric minimum.
Using (\ref{mass}), we find that the mass in the late time Universe is constrained by
\begin{equation}
m_{\rho 0} \gg m_{3/2}.
\end{equation}
The gravitino mass is typically always greater than $1$ eV in most scenarios such as gravity and gauge mediated supersymmetry breakings, see
\cite{Nilles:1983ge} and references therein. The mass of the dark energy field is therefore always very large compared to the Hubble rate now.

Supersymmetry breaking also introduces a shift in the vacuum energy density
\begin{equation}
\rho_\Lambda= \vert F\vert^2 - 3 m_{3/2}^2 m_{\rm Pl}^2
\end{equation}
where $F$ is the order parameter of supersymmetry breaking. This displacement
must be almost vanishing. In the following, we will find that a contribution from supersymmetry breaking to the
vacuum energy must be included in the supersymmetron model in order to comply with current data.

\section{Supersymmetron Cosmology}

\subsection{Cosmological dynamics}

Let us now discuss the dynamics of the model. In the radiation and matter eras, the field tracks the minimum (\ref{min})
with an energy density
\begin{equation}
\frac{V_F(\phi_\rho)}{\rho_\psi}\approx \frac{x}{n} (\frac{\rho_\infty}{\rho_\psi})^{1/(n+1)}.
\end{equation}
At the minimum (\ref{min}), we also have
\begin{equation}
\frac{g\phi_\rho}{m} \rho_\psi \approx n V_F(\phi_\rho)
\end{equation}
implying that
\begin{equation}
V_{\rm eff}( \phi_\rho)\approx (n+1) V_F(\phi_\rho)
\end{equation}
 When the density reaches a value close to $\rho_\infty$, the previous expression is not valid anymore and
\begin{equation}
 V_F=\frac{1}{2} \frac{g^2 \rho_\psi^2}{m^2 m_0^2}\approx \frac{x}{n} \frac{\rho_\psi^2}{\rho_\infty}, \ \  V_{\rm eff}=\frac{g\phi_{min}}{m}\rho_\psi -\frac{1}{2} \frac{g^2 \rho_\psi^2}{2m^2 m_0^2} \approx x\rho_\psi(1-\frac{1}{n} \frac{\rho_\psi}{\rho_\infty})
\end{equation}
when the
field is close to the supersymmetric minimum.
The transition between the two regimes occurs when $\rho\approx \rho_\infty$.
As the mass of the scalar field is always greater then the Hubble rate, the minimum of the effective potential is always an attractor.

\subsection{Convergence to the attractor}

 The convergence of the supersymmetron field to the effective minimum is analogous to that of the chameleon field, which was studied in an
expanding universe in \cite{Brax:2004qh}. It was shown that the variance averaged over many oscillations is
\begin{equation}
<(\varphi-\varphi_\rho)>^2\propto \frac{\rho_\psi}{m_\rho}
\end{equation}
such that
the energy stored in the oscillatory motion
\begin{equation}
\rho_{\rm osc}= m^2_\rho <(\varphi-\varphi_\rho)^2>
\end{equation}
satisfies
\begin{equation}
\frac{\rho_{\rm osc}a^3}{m_\rho}= {\rm constant}.
\end{equation}
The energy density $\rho_{\rm osc}$ is therefore redshifted like
\begin{equation}
\rho_{\rm osc} \sim a^{-3- 3(1+w)/2 -3\beta/(n+1)}
\end{equation}
in the radiation and matter eras. The oscillatory energy density
decreases faster than the CDM density, hence it cannot be considered to be a new form of Cold Dark Matter and is, from early times. irrelevant to the cosmological dynamics.

\subsection{Supersymmetron and acceleration}

When the supersymmetron field coincides with the attracting minimum and upon using $A(\phi_\rho) \rho_\psi= \rho_\psi + nV_F(\phi_\rho)$, acceleration starts when
\begin{equation}
\rho_{\rm acc}= (\frac{x}{n}(2-n))^{n+1} \rho_\infty
\end{equation}
as long as $n\le 2$. This
corresponds to a redshift
\begin{equation}
1+z_{\rm acc}= (\frac{x}{n}(2-n))^{(n+1)/3} (1+ z_\star)
\end{equation}
where $z_\star$ is the redshift when the dark energy field converges towards the supersymmetric minimum i.e. $\phi_\rho \approx \phi_\infty$.
Along this attracting trajectory the equation of state is constant and reads
\begin{equation}
w_{\phi} \approx -\frac{1}{1+n}
\end{equation}
where we retrieve the fact that acceleration for $w_{\phi}<-1/3$ requires $n<2$.

At the time that
dark energy reaches the vicinity of the supersymmetric minimum, the equation of state drops to a very small value
\begin{equation}
w_{\phi}\approx -\frac{1}{n}\frac{\rho_\psi}{\rho_\infty}\frac{1}{1-\frac{\rho_\psi}{n \rho_\infty}}
\end{equation}
Soon after converging to the vicinity of the supersymmetric minimum,  acceleration stops.
As long as $z_\star<1<z_{\rm acc}$, the universe is accelerating now. Acceleration started not so long in the past and will stop in the near future. It is a transient phenomenon only.
It is important to check that the mass of the dark energy field is large in order to evade gravitational tests. This is guaranteed as $m_\rho \gg m_{3/2}$. Hence dark energy does not lead to a long range fifth force here.

\subsection{Adding a cosmological constant}

In the recent past of the Universe and as long as $\rho_\psi > \rho_\infty$ we have
\begin{equation}
\rho_\phi \approx V_{\rm eff}(\phi_\rho) \approx \frac{x}{n}(n+1) \rho_\psi (\frac{\rho_\infty}{\rho_\psi})^{1/(n+1)}.
\end{equation}
This implies that the effective density fraction of the dark energy fluid is
\begin{equation}
\Omega_\phi\approx \frac{x}{n}(n+1) \Omega_\psi (\frac{\rho_\infty}{\rho_\psi})^{1/(n+1)}
\end{equation}
where $\Omega_\psi= \frac{\rho_\psi}{3 H^2 m_{\rm pl}^2}$.
This allows us to evaluate the acceleration parameter
\begin{equation}
q=\frac{1}{2}\frac{1+\frac{x}{n}(n-2)(\frac{\rho_\infty}{\rho_\psi})^{1/(n+1)}}{1+(n+1)\frac{x}{n} (\frac{\rho_\infty}{\rho_\psi})^{1/(n+1)}}.
\end{equation}
This is marginally consistent with data when $n\ll 2$ and $x\sim n$.
As the Universe is observed to be flat,  implying that $\Omega^0_\phi +\Omega^\psi_0=1$, we find that  for $x\sim 0$ and $\rho_\infty \sim \rho_\psi^0$ we have that
\begin{equation}
\Omega_\psi^0\sim \Omega_\phi^0 \sim 0.5
\end{equation}
where $\Omega_\psi^0$ and $\Omega_\phi^0$ are the CDM and DE fractions now. This
is marginally incompatible with data. Moreover, one must impose that the index $n$ must be unnaturally small. Hence it seems that a non-zero cosmological constant must be introduced. This is similar to the case of the symmetron
model \cite{Hinterbichler:2010es}, but in our case the cosmological constant
could be the result of supersymmetry breaking.

Assuming that supersymmetry breaking provides such a $\rho_\Lambda$, the equation of state of dark energy becomes
\begin{equation}
w_{\rm DE}= \frac{p_\phi -\rho_\Lambda}{\rho_\phi +\rho_\Lambda}
\end{equation}
which can be evaluated now when
\begin{equation}
\Omega_\Lambda+ \Omega_\psi^0 +\Omega_\phi^0=1.
\end{equation}
This reads
\begin{equation}
\Omega_\Lambda + \Omega_\psi^0 + (n+1)\frac{x}{n}\Omega_\psi^0(\frac{\rho_\infty}{\rho_\psi^0})^{1/(n+1)}=1
\end{equation}
leading to the dark energy equation of state
\begin{equation}
w_{\rm DE}= -1 + x \frac{\Omega_\psi^0}{1-\Omega_\psi^0} (\frac{\rho_\infty}{\rho_\psi^0})^{1/(n+1)}.
\end{equation}
This is close to -1 when $x$ is small.
Moreover the acceleration parameter becomes now
\begin{equation}
q_0=-1 +\frac{3(x+1)\Omega_\psi^0}{2}
\end{equation}
which implies that we find an agreement with data for a small value of $x$ and  a non-vanishing cosmological constant.

\subsection{Growth of structures}

The effect on the growth of structure of the supersymmetron are potentially very interesting. Indeed, the coupling to matter is given by
\begin{equation}
\beta_\phi= m_{\rm Pl} \frac{d\ln A}{d\varphi}.
\end{equation}
This is explicitly
\begin{equation}
\beta_\phi= \frac{g m_{\rm Pl}}{m} \frac{1}{k(\phi_\rho)} \frac{1}{1+ x(\frac{ \rho_\infty}{\rho_\psi})^{1/(n+1)}}
\end{equation}
in the matter era.
Structures,  represented by the CDM density contrast $\delta$,  grow in conformal time according to
\begin{equation}
\delta'' + {\cal H} \delta' -\frac{3}{2} {\cal H}^2 (1+ 2\beta^2_{\rm eff})\delta=0.
\end{equation}
Structure formation is affected when
\begin{equation}
\beta^2_{\rm eff} \equiv \frac{\beta_\phi^2}{1+ \frac{m_\rho^2 a^2}{k^2}}
\end{equation}
is of order one. Now as astrophysical scales are much larger than particle physics scales we have
\begin{equation}
\beta^2_{\rm eff}= \frac{\beta_\phi^2 k^2}{a^2 m^2_\rho}.
\end{equation}
Before the convergence to the supersymmetric vacuum, this reduces to
\begin{equation}
\beta_{\rm eff}^2= \frac{2}{n} \frac{m_{\rm pl}^2}{\rho_\infty}\frac{1}{(1+ x(\frac{ \rho_\infty}{\rho_\psi})^{1/(n+1)})^2}(\frac{\rho_\infty}{\rho_\psi})^{1+ (2\beta-1)/(n+1)}\frac{k^2}{a^2}.
\end{equation}
The effect of modified gravity is significant when
\begin{equation}
\frac{k}{a}\gtrsim \sqrt{\frac{n}{2}} (\frac{\rho_\psi}{\rho_\infty})^{1/2+ (2\beta-1)/2(n+1)}(1+ x(\frac{ \rho_\infty}{\rho_\psi})^{1/(n+1)}) \frac{\sqrt{\rho_\infty}}{m_{\rm Pl}}.
\end{equation}
Now as $\rho_\infty \lesssim \Omega_m^0\rho_c$, we find that gravity is modified on scales smaller than
\begin{equation}
\frac{k_{\rm mod}}{a} \approx \sqrt{\frac{3n}{2}} (\frac{\rho_\psi}{\rho_\infty})^{1/2+ (2\beta-1)/2(n+1)}(1+ x(\frac{ \rho_\infty}{\rho_\psi})^{1/(n+1)}) H_0.
\end{equation}
This scale reaches a value of order $H_0$. In the matter era and the acceleration epoch, this scale is of astrophysical order and could lead to interesting effects on the growth of structures. This is left for future investigation.

\section{Conclusion}

We have considered a model of supersymmetric dark energy. We have shown that dark energy can appear as a small supersymmetry breaking effect due to the non-vanishing of the CDM number density. One of the prominent features of this model is the fact that the acceleration of the universe is not eternal but only lasts a finite amount of time. We have also made sure that the supersymmetry breaking corrections have a negligible influence on the dynamics of the model. This guarantees that the supersymmetron has a mass much larger than the gravitino mass thus evading all gravitational tests.
In this scenario, the future of the universe would be a nearly supersymmetric vacuum, only broken by the decreasing matter density. Acceleration would be  a transient phenomenon. Unfortunately, this scenario fails to reproduce the apparent fraction in dark energy now which is close to three quarters. We have therefore been compelled to reintroduce a  contribution to the vacuum energy. This energy density could result from supersymmetry breaking effects. The necessity to have such a cosmological constant is a nuisance although we hope that improved models will allow one to have the right amount of dark energy from the
cosmological breaking of supersymmetry.
Even with this extra cosmological constant, the supersymmetron model has potentially interesting effects on the growth of cosmological structures which are under scrutiny.

\acknowledgments
A.C.D. is supported in part by STFC and wishes to thank IPhT Saclay for hospitality whilst this work was in progress.
We would like to thank C. van de Bruck and J. Martin for relevant suggestions and H. Winther for numerous observations and collaboration on a follow up paper.

\end{document}